\documentclass[a4paper,11pt]{article}
\pdfoutput=1 

\usepackage{jheppub} 

\usepackage[T1]{fontenc} 
\title{\boldmath Timing-based mass measurement of exotic long-lived particles at the FCC-ee}

\author[a]{R. Aleksan,}
\author[b]{E Perez,}
\author[c]{G Polesello}
\author[c,1]{N. Valle\note{Corresponding author.}}

\affiliation[a]{IRFU, CEA, Universit\'e Paris-Saclay, 91191 Gif-sur-Yvette cedex, France}
\affiliation[b]{CERN, Geneva, Switzerland}
\affiliation[c]{INFN Sezione di Pavia, Via Bassi 6, 27100 Pavia, Italy}

\emailAdd{roy.aleksan@cea.fr}
\emailAdd{emmanuel.perez@cern.ch}
\emailAdd{giacomo.polesello@cern.ch}
\emailAdd{nicolo.valle@cern.ch}

\abstract{
The very high luminosity run foreseen at the $Z$-pole for the FCC-ee will allow 
the detection in $Z$ decays of new particles with very low couplings to 
the Standard Model. These particles can have measurable flight paths before
they decay. If the  timing and 
the position of the decay vertex can be measured with high precision, 
the mass of such particles can be measured by exploiting the constrained
kinematics of an $e^+e^-$ collider. The mass resolution achievable with 
this technique is studied through a detailed analysis in the framework 
of a parametrised simulation of the performance of the IDEA detector.
The adopted benchmark model is the production of Heavy Neutral Leptons, 
which is one of the key channels for new physics discovery at the FCC-ee.}

\begin{document} 
\maketitle
\flushbottom

\section{Introduction}\label{sec:intro}
The next generation of proposed $e^+e^-$ circular colliders, such
as the CERN FCC-ee \cite{FCC:2018evy} will provide access to
a broad range of physics studies, from precision measurements
of the Higgs boson and of Standard Model (SM) parameters, to direct
searches for physics beyond the Standard Model (BSM).

Among the most promising BSM signatures at FCC-ee are long-lived particles
(LLPs), i.e. particles which decay inside the detector with a
measurable path. The detection of such particles poses severe 
performance constraints on the detector design, and a recent review 
of studies in this sector can be found in \cite{Blondel:2022qqo}.

As mentioned in \cite{Blondel:2022qqo}, the mass of LLPs can be measured by combining the length and 
    time of their decay in the detector with the kinematic beam constraints of FCC-ee.
    This possibility has been recognized for a considerable time, but the need for a large tracking volume, as well as
     a timing resolution in the ballpark of a few tens of picoseconds has so far placed the
     practical realisation of this idea out of experimental reach. These features are present
     in the proposed design for FCC potential detectors, and the present study formalizes and quantifies 
     for the first time the expected performance of this measurement technique in the FCC-ee context.

The production of heavy neutral leptons (HNL) in the decay of the 
$Z$ boson is used as a benchmark model,  and the constraints on the
experimental design are assessed on the basis of a parameterised 
simulation of the expected performance of the tracking system
of the proposed IDEA detector.
 
The paper is organised as follows: the basic analytical formulas 
are first introduced and discussed. In the following section 
the benchmark physics model and the detector simulation setup are
described. Finally the expected resolution on position and
timing of the vertex are evaluated in this framework, yielding an assessment
of the HNL mass measurement capabilites of the technique
in the parameter space of the HNL model considered.

\section{Mass measurement with vertex timing}\label{sec:formulas}
Consider a long-lived neutral particle $L$ of mass $M_L$ which is produced at 
an $e^+e^-$ collider with centre-of-mass energy $E_{CM}$ recoiling against 
a known particle of mass $M$, and decays into $\ge 2$ charged particles
at a distance $\delta d_L$ from the interaction point, with a delay 
$\delta t_L$  from the time of the interaction.
The momentum of $L$,  $p_L$, is expressed 
in special relativity as:
\begin{equation}
p_L=M_L\gamma\beta
\end{equation}
with $\beta=\delta d_L/\delta t_L/c$, with $c$ the speed of light,
and is constrained by the collision kinematics through the recoil relation:
\begin{equation}
p_L = \frac{ \sqrt{ ( E_{CM}^2 - (M_L+M)^2 ) ( E_{CM}^2 - (M_L - M)^2 ) } }{2E_{CM}}
\end{equation}
By eliminating $p_L$, $M_L$ can be measured as a function
of $\beta$, $M$ and $E_{CM}$, which are measurable or known 
quantities. 

This study explores what kind of precision
one can realistically obtain within the detector systems being
designed for experimentation at FCC-ee. To this purpose we work 
out analytically how the mass measurement depends on the 
detector performance,  for the case $M=0$,
corresponding to a LLP recoiling against a massless particle
such as a neutrino.  
In this case the formula for $M_L$ reduces to:
\begin{equation}
\label{eq:mass}
	M_L=E_{CM}\sqrt{\frac{1-\beta}{1+\beta}}\equiv E_{CM}F(\beta)
\end{equation}
Thus the mass of the long-lived particle can be extracted from 
the measurement of its $\beta$.

Before discussing the detector simulation, it is informative to
consider the expected mass resolution 
for values of $M_L$, $\delta d_L$ 
relevant to experimentation at FCC-ee.

To this end, the uncertainty in $M_L$  is related to the uncertainty on $\beta$ by the 
approximate formula:
\begin{equation}
	\sigma(M_L)=E_{CM}\frac{1}{\sqrt{\frac{1-\beta}{1+\beta}}(1+\beta)^2}\sigma(\beta)\equiv E_{CM}F'(\beta)\sigma(\beta)
\end{equation}
which is valid for sufficiently high values of $m_L$ and $\delta d_L$, such that 
$\beta$ is significantly smaller than one.

\begin{figure}[!h]
\centering 
\resizebox{.49\columnwidth}{!}{%
\includegraphics{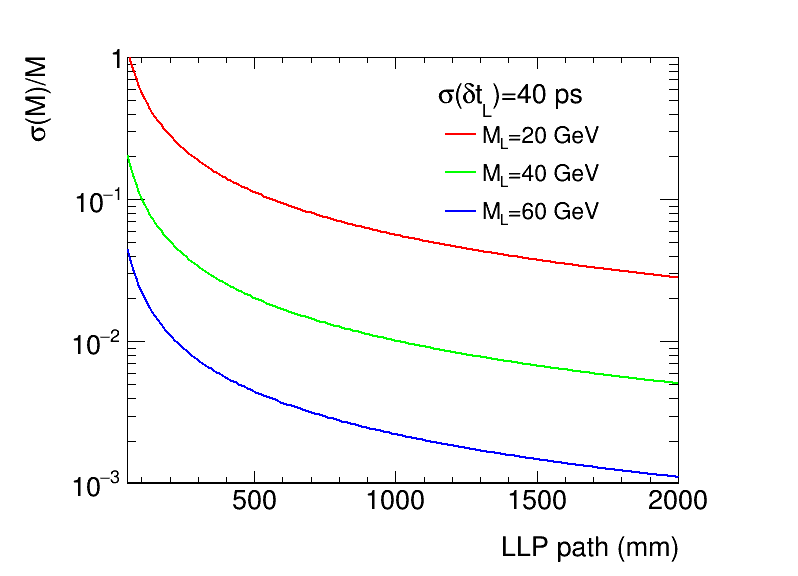}
}
\resizebox{.49\columnwidth}{!}{%
\includegraphics{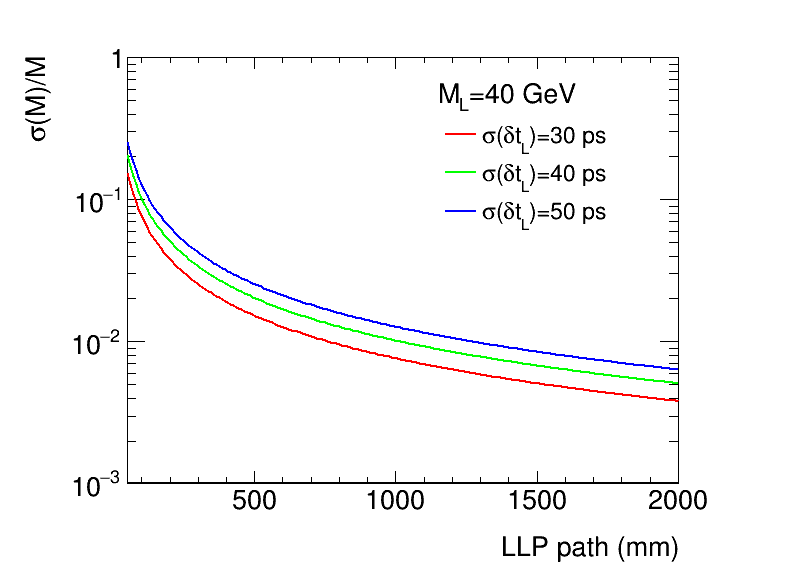}
}
\caption{\label{fig:dmvspath} Resolution of mass measurement through timing
as a function of the LLP decay path. Left: for different values of $M_{L}$ and an assumed uncertainty on the timing of 40~ps. Right: for $M_{L}=40~\mathrm{GeV}$
and three different values of the uncertainty on timing.}
\end{figure}

If the dominant uncertainty in the measurement of $\beta$ is the 
uncertainty on the $\delta t_L$, $\sigma(\delta t_L)$, this can be rewritten as
\begin{equation}
\label{eq:sigma1}
	\sigma(M_L)=\frac{E_{CM}\beta^2F'(\beta)c\sigma(\delta t_L)}{\delta d_L}
\end{equation}
The uncertainty on $M_L$ as a function of the decay path 
is shown in Figure~\ref{fig:dmvspath} respectively on the 
top for different values of $M_{L}$ and an assumed uncertainty
on the timing of 40~ps, and on the bottom for $M_{L}=40~\mathrm{GeV}$
and three different values of the uncertainty on timing.
For long lifetimes, above $\sim1~\mathrm{m}$, mass measurements 
below the percent level can be achieved, 
and the precision
strongly depends on $M_{L}$, being worse for lower values of
the mass.

\begin{figure}[!h]
\centering 
\resizebox{.49\columnwidth}{!}{%
\includegraphics{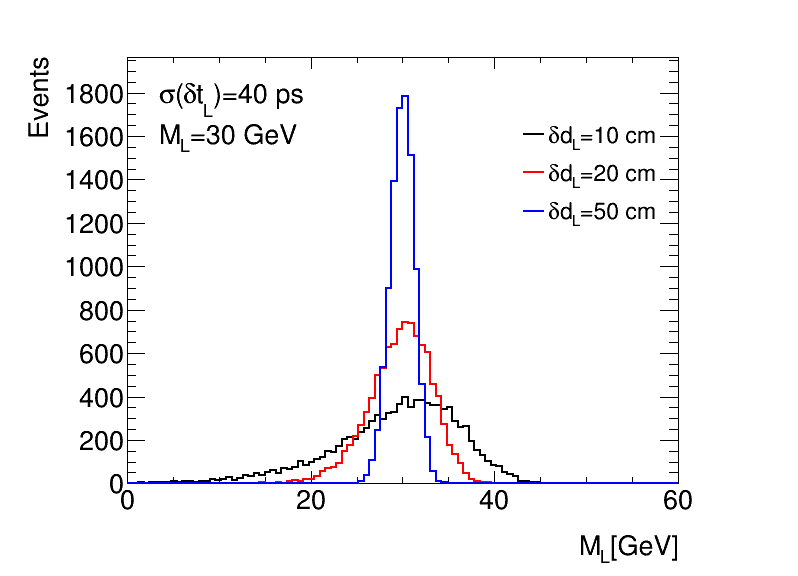}
}
\resizebox{.49\columnwidth}{!}{%
\includegraphics{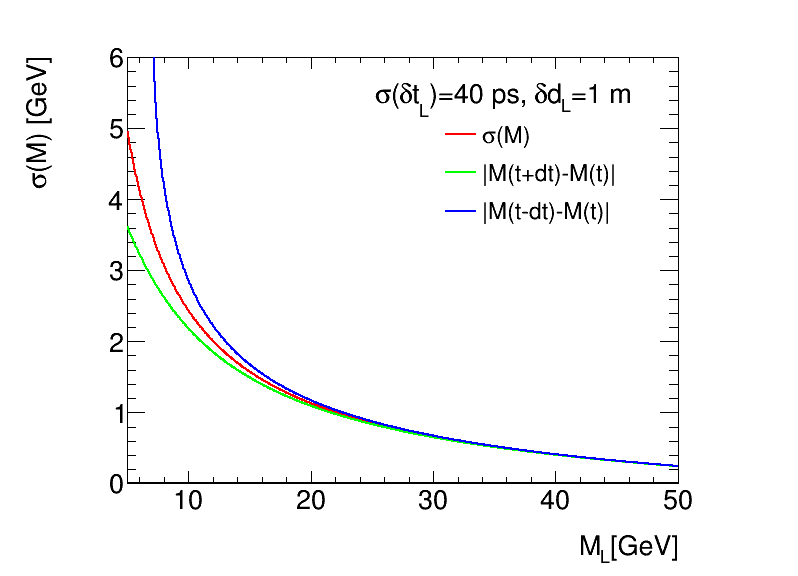}
}
\caption{\label{fig:asym} Left: distribution of the measured mass 
for a LLP of mass 30 GeV assuming a timing resolution of 40~ps for 
three different values of the LLP decay path: 10,20, and 50~cm. Right:
Mass resolution as a function of $M_L$ as calculated with the error 
propagation (red line) compared to the offset on mass measurement 
when the value of the timing is moved by $\pm1\sigma$ from the 
nominal value.}
\end{figure}

One should note that the distribution of the measured mass
is not Gaussian, but skewed towards low values, and the 
amount of asymmetry is a function of $\delta d_L$ and of the mass,
as shown on the top panel of Figure~\ref{fig:asym},
which depicts the distribution of the measured mass 
for a LLP of mass 30 GeV assuming a timing resolution of 40~ps for
three different values of the LLP decay path: 10, 20, and 50~cm, 
obtained with a toy Monte Carlo generation. The dependence
of this effect on the mass is shown on the bottom of Figure~\ref{fig:asym},
where the dependence of mass resolution from $M_L$, from equation~\ref{eq:sigma1}
is compared to the offset on mass measurement
when the value of the timing is moved by $\pm1\sigma$ from the 
nominal value for $\delta d_L=1$~m and $\sigma(\delta t_L)=40$~ps. 
The Gaussian approximation is good for $M_L>20$~GeV, but for $M_L$ 
below 10~GeV the low-mass tails become dominant. 

\begin{figure}[!h]
\centering 
\resizebox{.5\columnwidth}{!}{%
\includegraphics{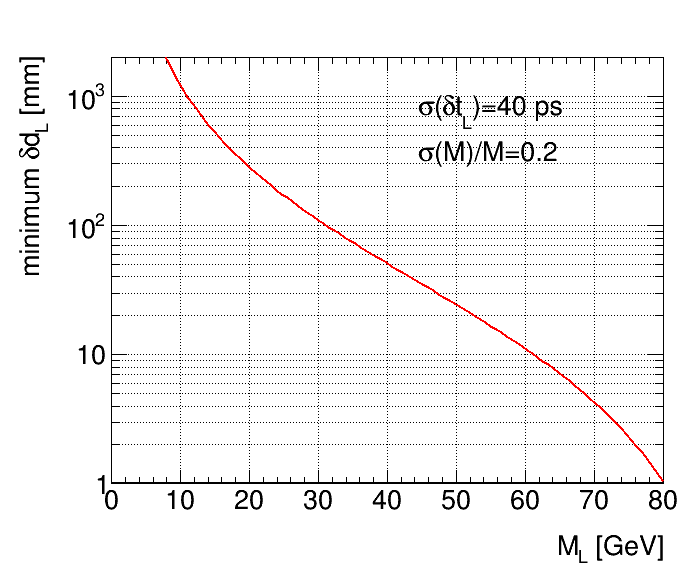}
}
\caption{\label{fig:mindh} Minimum decay length needed for achieving 
a 20\% error on the mass measurement as a function of the mass of the
LLP. The assumed resolution on timing is 40~ps.}
\end{figure}

An additional interesting quantity is the minimum path length necessary 
for achieving a given value of $\sigma(M_L)$ as a function
of $M_L$. This quantity  determines 
the available statistics of measurable events as a function
of the LLP lifetime. This is shown in Figure~\ref{fig:mindh}
for $\sigma(\delta t_L)=40$~ps, using as a benchmark value a 20\% resolution 
on the mass measurement. At low masses, below 20 GeV, only decays 
of the order of one meter yield a reasonable mass measurement, whereas 
for higher masses decays of a few centimetres are needed.

\section{The simulation setup}\label{sec:simulation}
In order to verify whether the technique described in 
Section~\ref{sec:formulas} can work on a realistic BSM model
we perform a study of the production of Heavy Neutral Leptons 
in the IDEA detector \cite{Antonello:2020tzq} at the FCC-ee. 
Generated signal events are passed through a parameterised simulation of the detector, 
and the parameter space of the model where the timing-based
measurement can be applied is assessed.
\subsection{Model definition and signal generation}
\begin{figure}[tbp]
\centering 
\resizebox{\columnwidth}{!}{%
\includegraphics{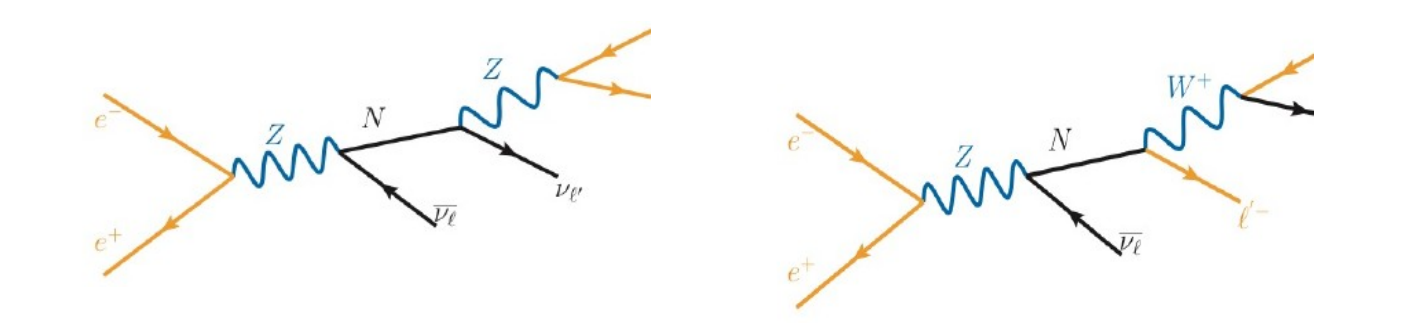}
}
\caption{\label{fig:HNL} Diagrams for production and decay of a heavy neutral 
lepton in the decay of a $Z$ boson.}
\end{figure}

The production of the HNL has been identified as one of the most promising new physics channels for FCC-ee at the Z pole in a seminal 
paper of 2014 \cite{Blondel:2014bra}, where details of the model can be found.
The production of HNL in the decay of the $Z$ bosons takes place through mixing 
with light neutrinos, yielding a HNL recoiling against a SM neutrino.
The HNL decays into three fermions,  through a virtual $W$ or $Z$ boson,
as shown in Figure~\ref{fig:HNL}.
If only one HNL flavour is assumed, mixing only with one SM neutrino,
the model is defined by two parameters: the HNL mass ($M_{HNL}$) and  the mixing parameter $U$ with the active neutrino. Long-lived (detached vertices) and prompt signatures are both possible depending on the the value of those parameters.

Several different decay channels and lifetime scenarios are accessible
at the FCC-ee. Detailed
discussions of previous experimental studies targeting FCC-ee
are contained in \cite{Blondel:2022qqo,Abdullahi2022jlv}.
As a benchmark for the present study the decay $HNL\rightarrow\mu jj$, 
is studied, which has a branching fraction of $\sim50\%$.
The simulated samples and the analysis code are the ones 
from the analysis documented in \cite{gpnv2023}, where more 
details can be found. 

The model of interest is implemented in the  
\verb+SM_HeavyN_LO+ \cite{Atre:2009rg,Alva:2014gxa,Degrande:2016aje} UFO, 
and signal samples were generated with \verb+MG5aMC@NLO+ \cite{Alwall:2014hca}.
The masses of the heavy neutrinos $N_{2}$ and $N_{3}$ were set to 10 TeV, and all of the 
mixing terms were set to zero, except for the mixing between the
muon and the HNL named as $N_{1}$ in the model. The associated production 
of a muon (anti)neutrino and the heavy neutrino $N_{1}$ was simulated,
with the $N_{1}$ directly decayed into a  muon and a quark-antiquark pair 
in the MG5 process card, so as to have the correct decay kinematics.

We performed a scan of the mass of $N_{1}$ between 5 and 85 GeV.
For each mass, coupling values were simulated in a range 
going from the minimal coupling yielding at least one event 
decaying within 2.5 meters from the centre of the detector for the
full FCC-ee Z-pole statistics, and a coupling squared of 
$5\times10^{-4}$ which is excluded by existing experiments.
The LHE files generated with \verb+MG5aMC@NLO+ are hadronised with 
\verb+PYTHIA8+ \cite{Sjostrand:2014zea} and then fed into 
the \verb+DELPHES+ \cite{deFavereau:2013fsa} fast simulation of the IDEA 
Detector, based on the official data cards used for the 
``Winter2023" production of backgrounds \cite{winter2023setup}.

For  the backgrounds from $Z$ decays, the official samples produced
by the central software group for the FCC PED studies under the tag
``Winter2023" were used \cite{winter2023samples}. 
The irreducible background from the four-fermion process
$
e^+e^-\rightarrow \mu \nu j j 
$
was produced at LO with \verb+MG5aMC@NLO+, and passed through 
the same \verb+PYTHIA8+-\verb+DELPHES+ chain as the signal events. 

\subsection{The IDEA detector and its simulation}\label{sec:idea}

The IDEA detector concept is a proposal for a general-purpose detector for the FCC-ee.
The design  includes
an inner detector composed of 5 Monolithic silicon 
pixel (MAPS) layers
followed by a high-transparency and  high-resolution drift chamber.  
We assume the presence of a timing layer positioned around the inner tracker at an approximate radius of 2 meters from the interaction point. 
A silicon wrapper employing Resistive Silicon Detector technology has been proposed for this purpose to enhance the drift chamber’s capabilities. This addition would effectively improve particle identification by providing precise spatial tracking along with timing measurements, achieving time resolutions on the order of tens of picoseconds, as documented in \cite{Cartiglia:2023izc}.
A superconducting solenoid producing a 2~T magnetic
field surrounds the inner detector.
Outside the solenoid is  a high-resolution  dual-readout 
fibre calorimeter, preceded by two preshower layers built with $\mu$-rwell technology,
and followed by three $\mu$-rwell  layers for muon detection, embedded in the 
return yoke of the solenoid.

The present analysis relies on the parameterised simulation in \verb+DELPHES+ of the 
inner detector for the estimate of the tracking and vertexing performance
of IDEA. 
The vertex detector is simulated as 5 cylindrical layers with 
radius between 1.2 and 31.5 cm with 2-D readout, and resolution $3~\mu$m
except the outermost layer with a resolution of $7~\mu$m,
and 3 disks in each of the forward/backward regions with a resolution of $7~\mu$m.
The drift chamber is modelled as 112 co-axial layers, arranged in 24 identical azimuthal sectors, at alternating-sign stereo angles ranging from 50 to 250 mrad, and
an assumed resolution on the single measurement of $100~\mu$m. The chamber has
a length of 4 meters, and covers the radius between 34~cm and 2~m, yielding 
full tracking efficiency for 
polar angles $\theta$ larger than about $10$ degrees.
The DELPHES simulation software relies on a full description of the geometry of the IDEA vertex detector and drift chamber and accounts for the finite detector resolution and for the multiple scattering in each tracker layer. It turns charged particles emitted within the angular acceptance of the tracker into five-parameter tracks (the helix parameters that describe the trajectory of the particle, including the transverse and longitudinal impact parameters), and determines the full covariance matrix of these parameters.
Vertexes are reconstructed using these tracks as 
input, based on a simple $\chi^2$ minimisation with constraints, yielding 3D
vertexes with their $\chi^2$ and covariance matrix. More details on the vertexing code used here can be found in ~\cite{Bedeschi_vertexing}.

\section{Analysis and results}
A full analysis based on the simulated samples for signal and
background described in the previous section was performed.
The analysis is focused on the run at the $Z$-pole, with an assumed 
statistics of $6\times10^{12}$ produced $Z$ bosons. 
The aim is to evaluate, for the benchmark model of HNL production,
if there is a sizeable chunk of the parameter space where 
the timing-based mass measurement technique can be applied,
and the expected resolution on the mass of the HNL. The detailed implementation of the analysis, and the resulting performance figures depend on the choice of the model. Therefore in the following the discussion of the analysis will refer to $M_{HNL}$  rather than to the generic $M_{L}$ introduced in Section~\ref{sec:formulas}.

\begin{figure}[tbp]
\centering 
\resizebox{0.5\columnwidth}{!}{%
\includegraphics{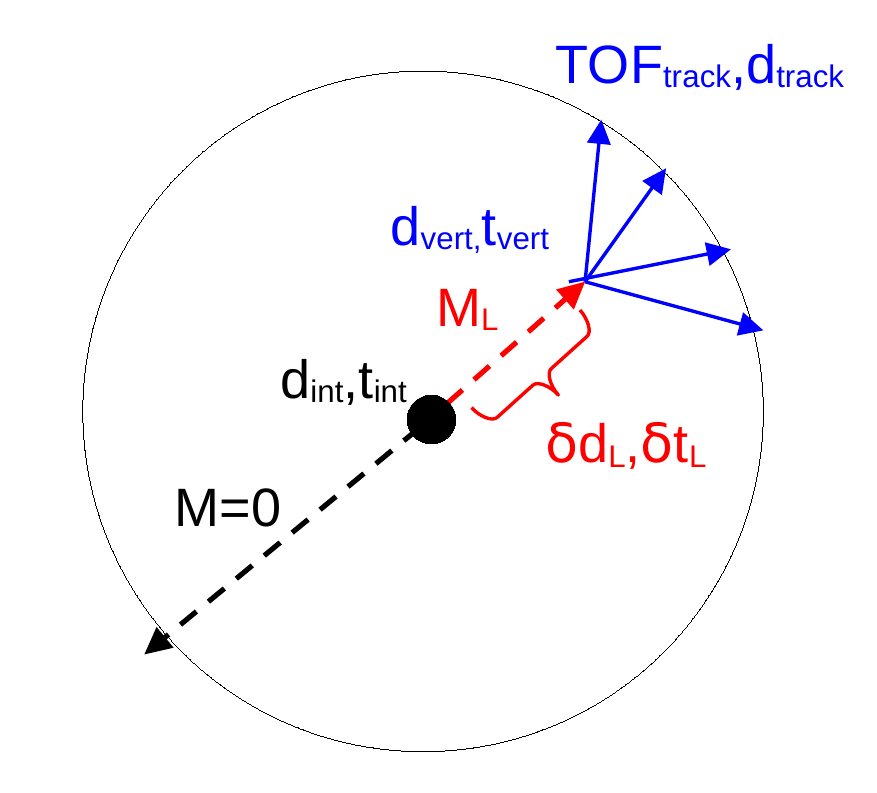}
}
\caption{Cartoon showing the different factors contributing to the
measurement of the vertex timing.
\label{fig:cartoon} }
\end{figure}

A cartoon showing the projection of the relevant event topology in 
the plane transverse to the beam is shown in Figure~\ref{fig:cartoon}.

As explained in Section~\ref{sec:formulas}, in order to calculate $M_{L}$
the quantities  $\delta d_L$ and $\delta t_L$ need to be evaluated, with:
\begin{equation}
\delta d_L\equiv ||\overrightarrow{OV}-\overrightarrow{OI}||
\end{equation}
and
\begin{equation}
\mbox{$\delta t_L\equiv t_{vert}-t_{int}$}.
\end{equation}

Here $\overrightarrow{OI}$ with magnitude $d_{int}$ is the vector connecting the interaction vertex
to the center of the detector, and $t_{int}$ the time difference between the interaction (primary) vertex
and the zero time for the beam crossing.
The equivalent quantities for  the LLP decay (secondary) vertex are
named $\overrightarrow{OV}$, $d_{vert}$ and $t_{vert}$

Among these quantities, the three components of $\overrightarrow{OV}$,  and hence their
magnitude $d_{vert}$ are measured directly in the tracking detector.

For each track attached to the decay vertex, three quantities can be measured:
$\mathrm{TOF}_{track}$, the arrival time of each charged particle to the timing 
layer situated around the outer radius of the IDEA tracker; the track momentum; 
and $d_{track}$, the length of the path of the track in the inner detector
before reaching the timing layer. 

The time $t_{vert}$ is not directly measured, but it can be calculated 
on the basis of $d_{vert}$ and of the measured track quantities
through an algorithm which will be described in the following.

The experimental uncertainties on $d_{vert}$, $d_{track}$ and momentum 
are determined by the design of the inner detector described above 
and are a function of the position of the
decay vertex and of the number of tracks attached to it.
The resolution on $\mathrm{TOF}_{track}$, 
($\sigma(\mathrm{TOF})$), does not depend on the topology
of the event, and is determined by the technology chosen for the implementation
of the tracking layer.

The quantities $t_{int}$ and $d_{int}$ cannot be measured, as the 
primary interaction results in two invisible neutrinos, and are constrained 
by the parameters of the accelerator.

The expected experimental uncertainties on the various quantities of
interest are discussed in the following sections. The results are then 
combined into the desired  estimate of the uncertainty of the HNL mass 
measurement. 

\subsection{Uncertainty on interaction point}

The $e^+e^-$ interaction point is centered in the center of the detector,
and its time is centered on the nominal beam crossing time, which 
are taken as values of $d_{int}$ and $t_{int}$ for the calculation 
of the decay path of the HNL.

The shape of the interaction point at the FCC-ee, can be approximated  
as a 4-dimensional Gaussian distribution,
determined by the size and shapes of the particle bunches in the
accelerator and by their crossing angle.

\begin{figure}[tbp]
\centering 
\resizebox{\columnwidth}{!}{%
\includegraphics{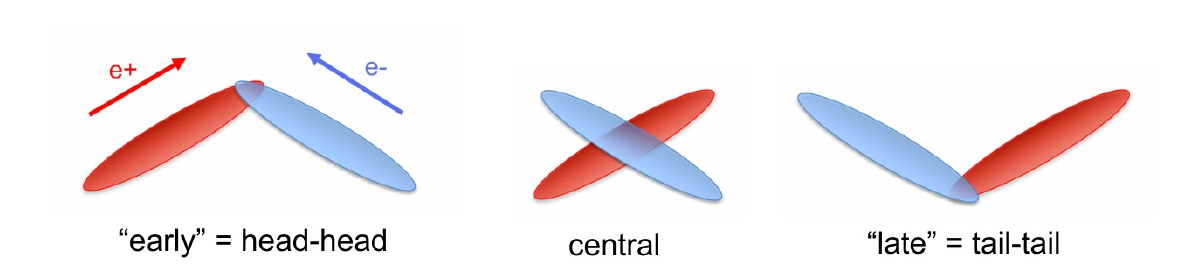}
}
\caption{Cartoon showing how the spread in time/position of the interaction vertex is generated in the crossing of the electron and positron beams.
\label{fig:beam_beam} }
\end{figure}

With the FCC-ee parameters used for the current
study~\cite{talk_IS_workshop_december,interactionVertex}, 
the interaction vertex will have a
spread
$\sigma_x=5.96~\mathrm{\mu m}$, 
$\sigma_y=23.8~\mathrm{nm}$, $\sigma_z=0.397~\mathrm{mm}$,
$\sigma_t=36.3~\mathrm{ps}$.
A cartoon showing how this spread arises is shown in Figure~\ref{fig:beam_beam}.

This  spread in time and position of the interaction point will thus 
contribute to the uncertainty on the measurement of the LLP 
path length in the detector.

An additional uncertainty is the jitter in the reference signal of beam crossing delivered by the machine to 
the experiment, and how reproducible the centering of the beam crossing is 
fill by fill. From the LHC experience 
\cite{Baron:2012fb}, the intrinsic jitter of the signal generated by RF of the accelerator is order 2~ps, with an additional 2~ps from the transmission to the experiments. As regards the reproducibility of the bunch structure and position,
it is expected that it will be very precisely 
measured using the large number of available $Z$ decays. This small uncertainty is ignored in the following calculations.

\subsection{Uncertainty on vertex timing}

The timing layer, located around the circle shown in 
Figure~\ref{fig:cartoon}, measures the arrival time of the charged particles from the HNL decay to the 
outer radius of the inner detector  ($\mathrm{TOF}_{track}$). 
For each track attached to the decay vertex the value of $t_{hit}$, which
is the estimate for the track of the value of $t_{vert}$ is calculated with
the following procedure:
\begin{itemize}
\item
The origin of the track (the HNL vertex) and its momentum are known.
\item
The helicoidal trajectory in the magnetic field is known. 
From its intersection to the timing layer, the flight distance $d_{track}$ of the 
track from its origin to the layer can be calculated.
\item
If the mass of the particle producing the track is known, from the momentum and the flight distance
the time of flight of the track from its origin to the timing layer can be calculated.
\item
Subtract this time of flight from $\mathrm{TOF}_{track}$. This gives $t_{hit}$.
\end{itemize}
If one of the decay products of the HNL is 
a muon, one can apply this procedure to the muon alone,
which is well identified by the muon detector. However, if additional particles
are produced, and their time of flight can be measured reliably, the 
precision of the timing measurement can be improved by a factor $\sqrt{N}$ where
$N$ is the number of tracks used for the measurement.

\begin{figure}[!h]
\centering 
\resizebox{0.49\columnwidth}{!}{%
\includegraphics{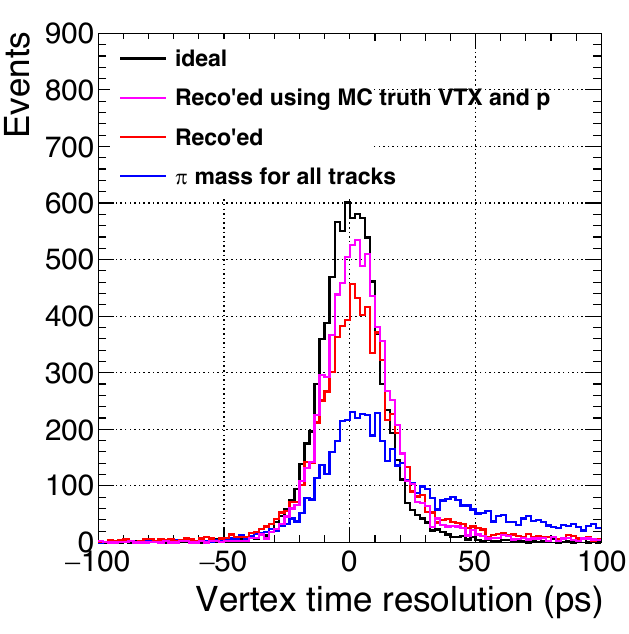}
}
\resizebox{0.49\columnwidth}{!}{%
\includegraphics{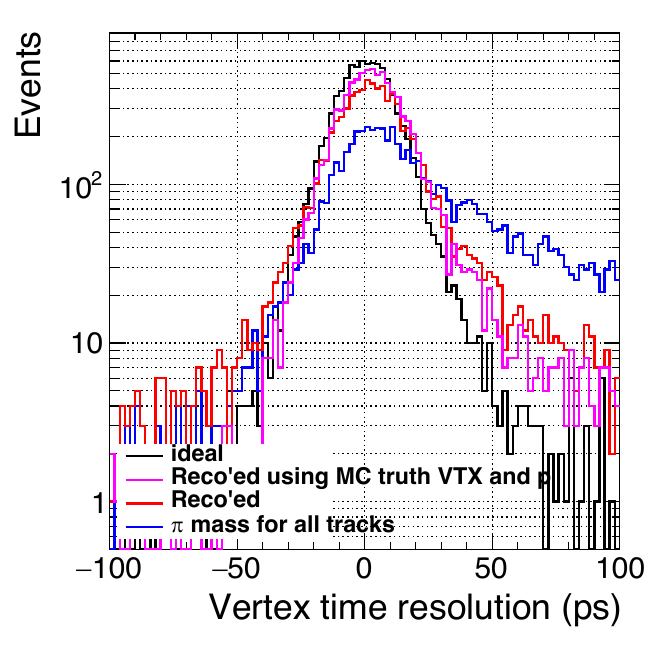}
}
	\caption{Distribution of measured vertex timing for $M_{HNL}=40$~GeV, $c\tau_{HNL}=1$~m, $\sigma(\mathrm{TOF})=30$~ps. The different lines correspond to different approximations for the measurement. s described in the text. Left: linear scale; Right: logarithmic scale to better appreciate the tails for wrong assignment of the masses of decay products.
\label{fig:tofplots} }
\end{figure}

The achievable resolution using all tracks was studied 
on a benchmark sample of $HNL\rightarrow\mu jj$ decays,
for $M_{HNL}=40$~GeV and proper time $c\tau_{HNL}=1$~m passed through the
parameterised simulation of the IDEA inner detector described in the previous section. 
The value of  $\mathrm{\sigma(TOF)}$ is assumed to be 30~ps, based on the detector design described in Section~\ref{sec:idea}.
If for each track the correct mass from Monte Carlo truth is assumed, the resolution on the timing is Gaussian, with an RMS of 11.2~ps, 
as shown in the black curves of the two distributions\linebreak shown in Figure~\ref{fig:tofplots}.
This is approximately consistent with a $1/\sqrt{\langle N_{tracks}\rangle}$ scaling of the resolution, 
as the average number of tracks attached to the vertex is \mbox{$\langle N_{tracks}\rangle=8.2$}.

A first approximation is to assign the pion mass to all of the tracks.
The resulting timing resolution is shown as a blue line in the plots in Figure~\ref{fig:tofplots}. A very large tail in the timing resolution appears, badly spoiling the measurement.

The IDEA drift chamber has been designed to have a particle identification
through cluster counting along a track, so it could be used
to choose a mass hypothesis. However,  for long HNL flight distances, 
the tracks are short, and it was found that the momentum of
tracks peaks is in the region where particle separation by ionisation is not 
very powerful.

Therefore for the present exploratory exercise, a simple iterative algorithm based 
only on the TOF of the particles was developed:
\begin{itemize}
\item
Use only tracks with $p_T>300$~MeV.
\item
First use the pion mass hypothesis for all tracks that are not muons or
electrons.
\item
For each hadron track, determine 
$\Delta t=| t_{hit} - Ave_{other}|$
where $Ave_{other}$ is the average of 
of all other tracks. Find
the track with the largest $\Delta t$.
\item
If $\Delta t>$ cut (set to $3\times36$~ps at the $Z$ peak), 
try instead the kaon and proton hypotheses and keep the one that minimizes $\Delta t$.
\item
Iterate: find again the track with the worst $\Delta t$ (the partial averages are
recomputed), excluding the one found previously, up to the point where no improvement on $\Delta t$ is possible.
\item The average of all values of $t_{hit}$ after the procedure is taken as the measurement of $t_{vert}$.
\end{itemize}
The result of this algorithm is shown as a red line in Figure~\ref{fig:tofplots}.
The tails are strongly reduced, although not fully eliminated. If a Gaussian fit
is applied, the resolution is only 25\% worse than in the ideal case. To understand 
the origin of the worsening in resolution, the algorithm was rerun using the `truth' 
Monte Carlo position of the HNL vertex and momenta of the tracks. The result is shown 
as a magenta line, and shows that a large part of the worsening is not from the algorithm,
but from the position and energy resolution of the detector. 

In Table~\ref{tab:sigmat} the values of the vertex timing resolution for $\sigma(\mathrm{TOF})=30$~ps
are shown for different values of $M_{HNL}$ and $c\tau_{HNL}$.
\begin{table}[!h]
\centering
\begin{tabular}{|l|c|c|c|c|c|}
\hline
	$M_{HNL}$ & $c\tau_{HNL}$ & Ideal & Reco & $\langle p\rangle$ & $\langle N_{tracks} \rangle$ \\
	(GeV) & (mm) & (ps)    & (ps)   & (GeV) &\\
\hline
	10 & 1000 & 16.8 & 20.4 & 6.25 & 4.3\\
	20 & 32 & 13.8 & 19.2 & 5.02 & 5.9\\
	20 & 100 & 13.8 & 18.6 & 4.95 & 6.1\\
	20 & 1000 & 14.4 & 21.0 & 4.87 & 5.9\\
	20 & 3200 & 13.8 & 19.8 & 4.85 & 5.8\\
	40 & 10 & 12.6 & 20.4 & 4.1 & 8.4\\
	40 & 100 & 12.0 & 18.0 & 4.1 & 8.5\\
	40 & 1000 & 12.6 & 22.8 & 4.02 & 8.2\\
	60 & 10 & 11.4 & 21.6 & 3.84 & 10.4\\
\hline
\end{tabular}
	\caption{\label{tab:sigmat} Vertex timing resolution for different values of 
$M_{HNL}$ and $c\tau_{HNL}$. Ideal is the resolution in ps obtained using the correct mass
	assignment for tracks from the vertex and Reco'ed is the resolution using
	the algorithm described in the text. The average momentum and multiplicity of tracks
	from the vertex are also shown.}
\end{table}

\subsection{Uncertainty on vertex position}

\begin{figure}[!h]
\centering 
\resizebox{.49\columnwidth}{!}{%
\includegraphics{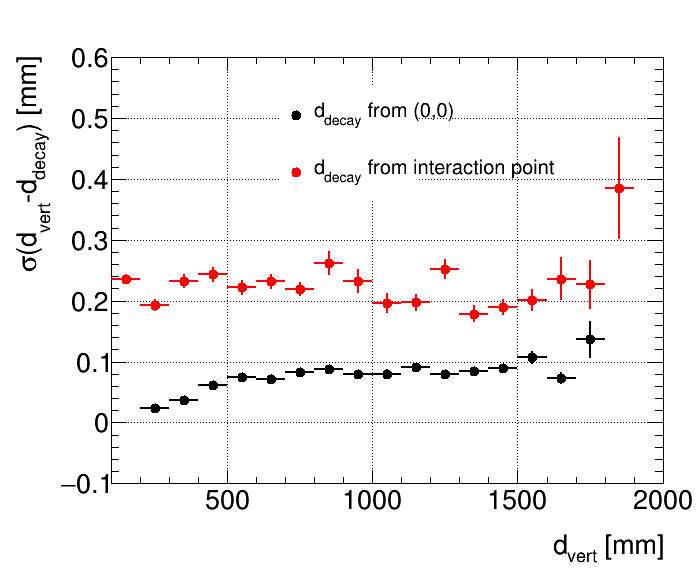}
}
\resizebox{.49\columnwidth}{!}{%
\includegraphics{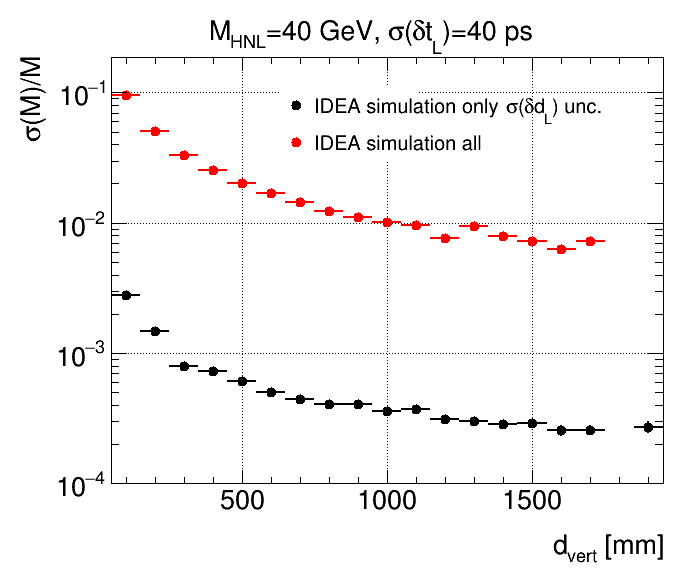}
}
\caption{\label{fig:vres} Left: Resolution on the position of the reconstructed vertex in the detector as a function of its position. Black points: intrinsic resolution of the vertex measurement. Red points: incorporate uncertainty on interaction point. 
Right: relative resolution on HNL mass measurement as a function of the position of the decay vertex. Black points: only uncertainty on vertex position considered. Red points:
both uncertainty on timing and position considered. The assumed value of $M_{HNL}$ is
40 GeV, and a total timing uncertainty of 40~ps is assumed.}
\end{figure}

Based on the parameterised simulation of the inner detector performance, 
the resolution of the measured distance of the vertex from 
to the centre of the detector $d_{vert}$ was evaluated 
for a sample of events with $M_{HNL}=40$~GeV as a function 
of the value of $d_{vert}$, and is shown as black points in the top 
panel of Figure~\ref{fig:vres}. The resolution is approximately at a value
of $\sim100~\mathrm{\mu m}$ when the vertex is fully in the drift chambers,
and it improves as the vertex approaches the beam pipe thanks to 
the higher resolution of vertex detector. The relevant quantity for
the mass measurement is however the HNL flight path, 
the distance from the unknown interaction
point. The sigma of the distribution of the difference between 
$d_{vert}$ and the HNL flight path is shown as red points 
in Figure~\ref{fig:vres} for the same HNL sample, and it 
is approximately flat over the whole measurement range at
a value between 200 and 250~$\mathrm{\mu m}$. 

In order to get a feeling of the relative importance of the uncertainty on position
and timing of the vertex on the measurement of mass, the mass resolution was
calculated assuming the timing perfectly measured (red points in the bottom panel
of Figure~\ref{fig:vres}), and based on analytical formulas
for a total timing resolution of 40~ps, incorporating both $\mathrm{\sigma(TOF)}$ and the 36.3~ps
uncertainty on the interaction time (black points). The impact of the timing resolution
is more than an order of magnitude larger than the one of position resolution.

\subsection{Uncertainty on mass as a function of flight path}

\begin{figure}[!h]
\centering 
\resizebox{.49\columnwidth}{!}{%
\includegraphics{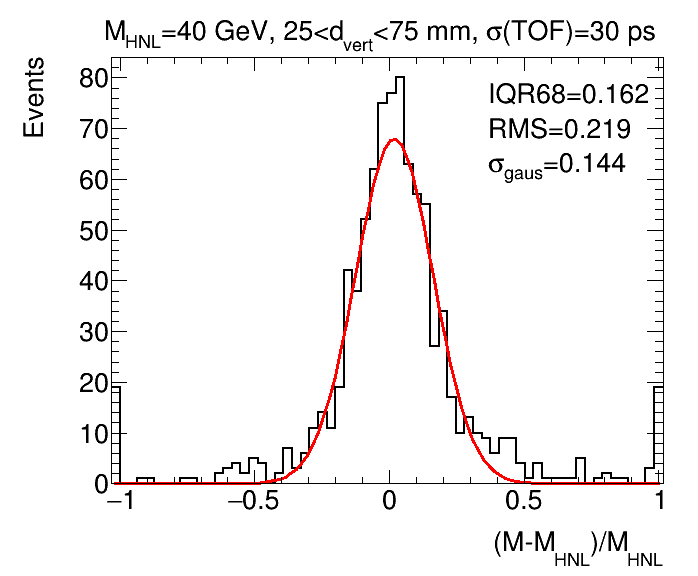}
}
\resizebox{.49\columnwidth}{!}{%
\includegraphics{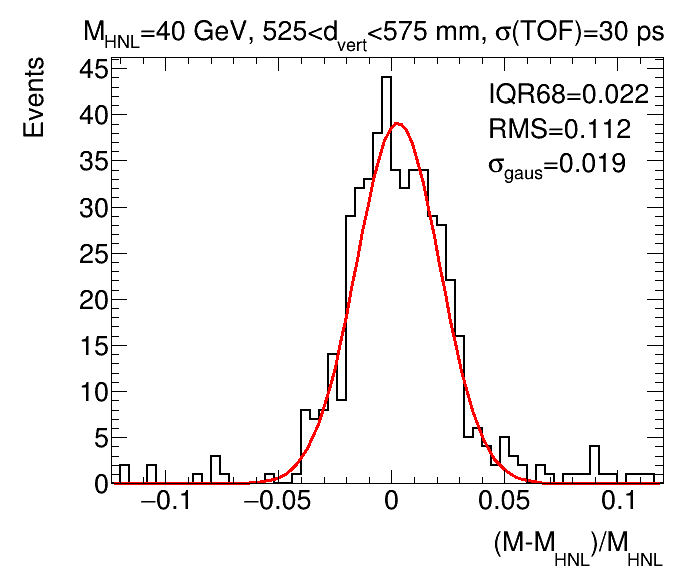}
}
\caption{\label{fig:massana} Relative  difference between the measured mass $M$ and the nominal HNL mass $m_{HNL}$ for two different bins in the vertex position $d_{vert}$. 
Assumed  TOF uncertainty of 30~ps and  $M_{HNL}=40~\mathrm{GeV}$.}
\end{figure}

Finally, the value of the HNL mass was calculated event-by-event 
on the basis of Formula~\ref{eq:mass}, where $\beta$ is calculated
using the timing of the vertex calculated with the TOF
measurement, and the distance $d_{vert}$ is calculated 
from the vertex position measured by the inner detector.
The two plots in Figure~\ref{fig:massana} show the distribution of the difference between the measured mass $M$ and  the nominal $M_{HNL}=40$~GeV scaled by $M_{HNL}$ for $\sigma(\mathrm{TOF})=30$~ps
and two bins in $d_{vert}$, centered around respectively 50 and 550~mm.
The two distributions show non-Gaussian tails on both sides, yielding a sigma
of the Gaussian fit significantly better than the RMS of the distribution.
As an estimator of the quality of the mass measurement the quantity IQR68 is 
defined, which is the interval between the quantile 16\% and the quantile 84\% of
the mass distribution.

\begin{figure}[!h]
\centering 
\resizebox{.49\columnwidth}{!}{%
\includegraphics{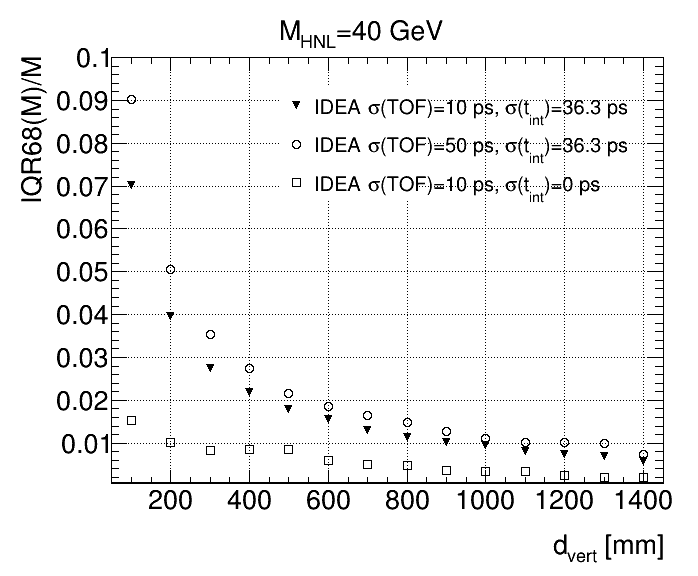}
}
\resizebox{.49\columnwidth}{!}{%
\includegraphics{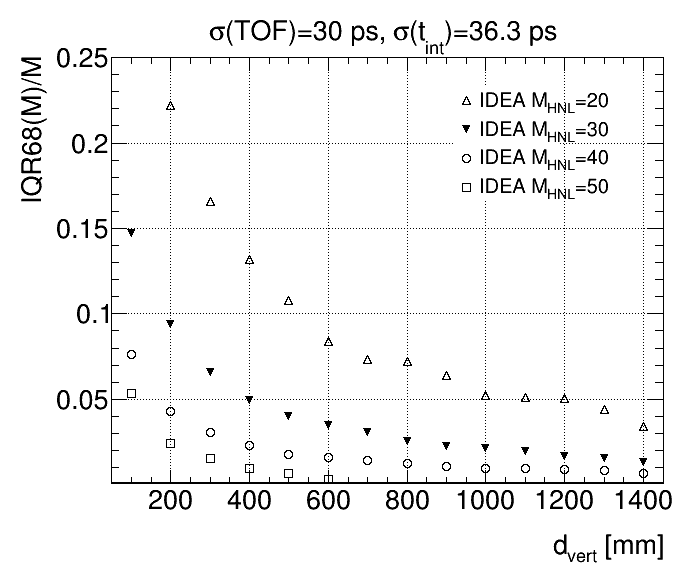}
}
\caption{\label{fig:ancomp} Resolution of mass measurement through timing
as a function of the position of the decay vertex. Left: $M_{HNL}=40$~GeV 
and three different configurations for timing resolution; Right: 
$\sigma(\mathrm{TOF})=30$~ps, and four values of $M_{HNL}$.}
\end{figure}

The distribution of the value IQR68/M as a function of $d_{vert}$ is shown in 
Figure~\ref{fig:ancomp}. 
The top panel shows the resolution for $M_{HNL}=40$~GeV, and two different 
values of $\sigma(\mathrm{TOF})$, 10 and 50~ps respectively. The resolution in the  50~ps case is only 20\% worse than in the 10~ps case, because the measurement is dominated by the lack of knowledge on the
time of the interaction vertex. 
It can be concluded that  value of 
$\sigma(\mathrm{TOF})$ starts having a significant impact on the measurement
starting from $\sigma(\mathrm{TOF})\sim 50$~ps.

Above $d_{vert}=200$~mm the HNL mass
is measured to better than 5\%, going below 1\% for $d_{vert}=1500$~mm.
As a comparison, the resolution is shown as well for the unphysical case where 
the timing of the interaction vertex is perfectly measured and $\sigma(\mathrm{TOF})=10$~ps. 
In this case the mass would be measured to better than 1\% starting from $d_{vert}=200$~mm.

In the bottom panel the same resolution is shown for $\sigma(\mathrm{TOF})=30$~ps and four
different values of $M_{HNL}$. As expected, the resolution improves sharply when 
$M_{HNL}$ increases.

\subsection{Parameter space coverage}

The achievable resolution on the mass as a function of the mass and of the path of a HNL
in the inner detector of IDEA has been evaluated in the previous sections.
The next step is the evaluation, in the framework of the benchmark model
described in Section ~\ref{sec:simulation}, and for the projected run
at the $Z$-pole at the FCC, of the parameter space for which
it is possible to measure the HNL mass with the timing technique described above.

For each model point the expected number of events $N_{sig}$ surviving the selections
in the $Z$-pole FCC run can be estimated. The study is performed on sets
of Monte Carlo events where the HNL decays semi-leptonically into a muon and two 
quarks, corresponding to approximately 50\% of the HNL decays.
The normalisation considers all of the visible decays of the HNL, corresponding
to approximately 93\% of the decays, as only a moderate dependence
of the timing measurement on the specific decay channel is expected. 

A  set of Monte Carlo experiments can be performed by selecting randomly in 
the sample of Monte Carlo signal events a number of events distributed according to 
a Poissonian distribution around $N_{sig}$. For each Monte Carlo experiment, 
the variance on the mass measurement strongly varies with the primary 
vertex position $d_{vert}$, and 
it is inversely proportional to the square of 
$d_{vert}$ 
as shown in Formula~\ref{eq:sigma1}.

The HNL mass is thus calculated for each experiment as the inverse variance weighted 
average of the measured masses, where the square of $d_{vert}$  is taken as weighting factor.
The uncertainty on the timing-based mass measurement is taken as the RMS of the
averages over a number of Monte Carlo experiments. 

The final state of interest is isolated by selecting events with
one and only one reconstructed muon, no other
reconstructed lepton, at least four reconstructed tracks and missing
momentum larger than 5~GeV.

In order to guarantee that the vertex position is well reconstructed,
the $\chi^2/ndf$ 
of the reconstructed vertex is required to be smaller than 10,
and that at most 5 of the reconstructed tracks in the event are not
used for the vertex reconstruction.  As discussed in \cite{gpnv2023}, this
selection strongly reduces the background from the tails of
$Z\rightarrow b\bar{b}$ production. We also require that the transverse
distance of the vertex from the centre of the detector is larger than 5~cm.

\begin{figure}[tbp]
\centering 
\resizebox{0.6\columnwidth}{!}{%
\includegraphics{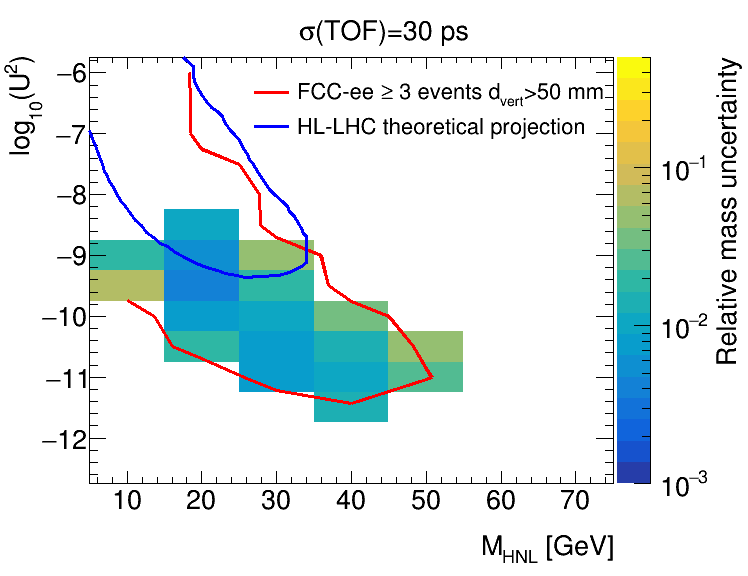}
}
\caption{\label{fig:reach} Expected relative resolution on mass measurement 
as defined in the text, in the \mbox{($M_{HNL}$, $U^2$)} plane for the
$Z$-pole run of the FCC-ee. The red line bounds
the region for which 3 events with $d_{vert}>50$~mm survive the cuts.
The blue line is a theoretical projection of the expected coverage
in the same plane of HL-LHC searches. The assumed value of
$\sigma(\mathrm{TOF})$ is 30~ps.}
\end{figure}

This last selection guarantees that the Standard Model backgrounds are reduced to 
a negligible level,
and that the mass measurement is not overly biased by the tails towards low masses
which appear when the decay path of the HNL is too short as discussed above.
The expected uncertainty on mass measurement obtained with this procedure
is shown in Figure~\ref{fig:reach} in the \mbox{($M_{HNL}$, $U^2$)} plane, for an assumed uncertainty on TOF measurement of
30~ps. No value is shown for model points for which $N_{sig}$ is larger than
the number of available Monte Carlo signal events. The red line bounds 
the region where $N_{sig}\geq 3$. The optimistic expected coverage of HL-LHC 
based on a theoretical estimate \cite{Drewes:2019fou} is shown as a blue line. 
Over all of the allowed region the HNL mass can be measured with the
vertex timing technique discussed in this paper with a precision of a few 
percent. This measurement is independent from the measurement which can be obtained
by measuring the visible mass in the detector using particle flow techniques 
combining information from the tracker and the calorimetric system. 
The final precision on the HNL mass will result from the combination of the two 
measurements.

\section{Conclusions}
The mass of a long-lived particle decaying within the FCC-ee detector can be determined by combining the timing and position of the decay vertex with constraints on beam energy and momentum.
Using the relevant analytic formulas, it was shown that for 
timing resolutions of a few tens of picoseconds, a resolution on 
the mass of few percent for LLP mass in the range 10-80~GeV can be expected.

On the basis of a parametrised simulation of the IDEA detector,
a complete analysis was performed for the production of an HNL in the process
$Z\rightarrow\nu\mathrm{HNL}$ and its decay into a muon and two jets.

For time-of-flight measurements of HNL decay products in the range of 10 to 50~ps, the mass resolution is dominated by uncertainties in the timing and position of the initial interaction, undetectable and constrained solely by the geometry of the beam-beam collisions.
For sufficiently long decay paths of the HNL, the analytical results 
are confirmed. A scan of the parameters space of the HNL model, assuming
the full statistics of the $Z$-pole run, shows that on a large part 
of the accessible space the experiments will be able to provide a measurement
of the HNL mass with a precision of a few percent, based on the 
proposed technique.

\acknowledgments
We would like to thank all of the participants in the FCC-ee study group for providing a stimulating working environment. Special thanks go to Franco Bedeschi for making his vertexing code available and for useful discussions about displaced vertices,
and to Patrick Janot for carefully reading the manuscript and useful comments.

The work of N.V. has received funding from the European Union’s Horizon 2020 Research and Innovation programme under  GA no 101004761.

\vspace{1em}

\noindent\textbf{Data Availability.} Part of the simulated Standard
Model samples are produced \linebreak
by the central software group for the FCC PED studies and are available at \url{https://fcc-physics-events.web.cern.ch/fcc-ee/delphes/winter2023/idea/}.

\vspace{1em}
\noindent\textbf{Code Availability.} The data cards for
the DELPHES fast simulation of the IDEA Detector and the official
FCC software are available at \url{https://github.com/HEP-FCC/}.

\end{document}